\documentclass[doublecol]{epl2}
\usepackage{amsmath}

\newcommand{\sq}{\unskip\nobreak\kern5pt\nobreak\vrule
     height4pt width4pt depth0pt}   
\title{Spectral action and neutrino mass}
\author{Andrzej Sitarz\thanks{Partially supported by MNII grant 189/6.PRUE/2007/7}}
\shortauthor{Andrzej Sitarz}
\institute{Institute of Physics, Jagiellonian University, Reymonta 4, 30-059 Krak\'ow, Poland}
\abstract{We propose the extension of the spectral action principle to
fermions and show that the neutrino mass terms appear then naturally
as next-order corrections.}

\begin{document}
\maketitle
\section{Introduction}

The problem whether neutrino is a massive or a massless particle,
which seemed to be one of long-standing puzzles of particle physics
seems to be solved by now. Convincing strong evidence of neutrino
oscillations, coming from various experiments (\cite{SK}, see
\cite{Alberico} for a review of further results) gave the necessary
experimental proof. Still, the subject is far from being closed for
theoretical and experimental physics as there remain open, at least,
three major questions: whether all neutrinos are indeed massive,
whether they are Majorana fermions and, probably one of the most
fascinating - still, if massive, why their masses are so small.

There exist, of course, various possible theoretical mechanisms,
which could be, in principle, verified experimentally (see for
instance \cite{Akhmedov} for a review). Here we would like to
address the issue from the point of view of geometry. In fact, none
of the fundamental symmetry principles seemed to enforce the
neutrino to be massless. Only recently (compared to the time-scale
of neutrino investigations) the early application of methods of
noncommutative geometry suggested that massless neutrinos agree with
the geometry of the Standard Model described as a finite spectral
triple or, in other words, a zero-dimensional manifold, satisfying
Poincar\'e duality in $K$-theory \cite{Connes,Connes07}.

The fact that the observed neutrinos are massive indicated that the model was
not very realistic and has led to some modified propositions, which admit
massive neutrinos and see-saw mechanism.
This, however, required some adjustments to the notion of geometry,
in particular the geometry of the finite space was required to
have a homological dimension different from $0$ (mod $8$).

There are probably many possible ways to adapt the model and its axioms to the physical reality,
or to interpret the physical data within the existing framework with slight changes of the particle
content of the model \cite{Schelp}.

Here, we shall discuss a possible mechanism, which relates the earlier explanations \cite{Weinberg}
with the current formalism of spectral geometry. This will have the advantage of providing not only
the explanation for the mass but also for its scale, while leaving the original geometry untouched.

\section{Standard Model and Geometry}

Within the framework of noncommutative geometry the standard Dirac operator has a natural extension by a
finite-dimensional matrix, which has a natural interpretation of the mass matrix and mixing angles - we
shall not discuss here in details the restrictions set on this matrix by the spectral triple axioms
(like the order-one condition). Let us briefly remind that for a four-dimension manifold, the Dirac operator
satisfies certain commutation rules with the chirality $\gamma$ and the charge conjugation $C$ operators,
which (we consider only even dimensions) are fixed. The difference between Minkowski and Euclidean
approach is in the mutual relations between the charge conjugation and the chiral projections: in
$1+3$ dimensions charge conjugation necessarily changes chirality, which is not the case in $0$ or $4$
dimensions.

The new notion of the Dirac operator allows for possible finite components of $D$, so that it becomes:
\begin{equation}
D = \gamma^\mu \partial_\mu + \omega_s + D_F,
\end{equation}
where $\omega_s$ is the spin connection and $D_F$ is a linear operator on the full Hilbert space.
The existence of this finite matrix component leads to the introduction of additional gauge fields
that could be naturally identified with the Higgs doublet.

However, what is crucial for us, is the assumption that the action function is purely
{\em spectral}, that is for the bosonic part it is taken as a cut-off trace of the Dirac
operator \cite{CoCha}:
\begin{equation}
S_b = \hbox{Tr} f(D^2), \label{boso}
\end{equation}
with suitable choice of the cutoff function $f$.

For the fermions one usually takes the expectation value of $D$:
\begin{equation}
S_f = <\Psi | D \Psi>, \label{ferm}
\end{equation}

The bosonic spectral action in its asymptotic expansion gives rise to all gravity-dependent terms,
which include the cosmological constant and the Einstein-Hilbert action as well as all terms
of the Yang-Mills-Higgs model, with the symmetry-breaking Higgs potential. The model is still open
for possible extensions, like models including leptoquarks \cite{PSS} or extended symmetries \cite{stephan}.

The mass terms for fermions arise in (\ref{ferm}) through the
nonzero expectation value of the Higgs from the minimal couplings
between Higgs and leptons, due to the symmetry-breaking potential term
for the Higgs doublet, which arises naturally in (\ref{boso})
\cite{Connes,IKS}.

Note that the spectral action works only in the Euclidean setup, so in order to pass to the
Minkowski setup one needs first to derive all local terms and then Wick rotate them.

We shall avoid this issue and also we shall not assume any
particular noncommutative geometry model, thus not touching the
fermion-doubling problem \cite{GraBo,Lizzi} and the need of reducing
the product of the Hilbert spaces to a certain physical subspace (as
in \cite{Connes07}). Instead, we take a minimalist approach and
consider only the spectral action of the already given part of the
Dirac operator from the physical Standard Model.

\section{Fermionic mass terms and neutrinos}

Majorana particles are fermions, which are invariant under the charge conjugation operation.
From the mathematical point of view, in the physically relevant Minkowski setup they are real
representations of the of the real Clifford algebra $\hbox{Cliff}(1,3)$. It is allowed that
Standard Model neutrinos are Majorana, with a left-handed neutrino and its right-handed
antipartner (in the Minkowski setup).  In principle, there are no obstacles that such particles
might be massive and one introduces a Majorana mass term of the form:

\begin{equation}
m_l \overline{\Psi_L^c} \Psi_L + h.c.,
\label{majmass}
\end{equation}

where $\Psi_L$ and $\Psi_L^c$ are chiral components of the
Majorana field $\Psi$.

It is important to realize that such terms are not excluded by
the Standard Model, however, to preserve the gauge invariance,
they would require the interactions of the type \cite{Weinberg}:

\begin{equation}
{\mathcal L}_{m} = \kappa \left(\overline{e_L^c} H^+
- \overline{\nu^c} H^0 \right) \left( H^+ e_L -  H^0 \nu \right),
\label{inter}
\end{equation}

where $e_l, \nu$ is the doublet of left-handed leptons, $H^+,H^0$
are the Higgs field components and $\kappa$ is a coefficient (or a
matrix if we take into account flavors). Such term, though
acceptable from the classical point of view, introduces, in the
context of quantum field theory, a nonrenormalizable interaction
\cite{Wilczek}. It is also clear that it cannot appear in the
standard noncommutative formulation of the fermionic action
(\ref{ferm}), since the coupling between the generalized Dirac
operator and the fermions is linear in $D$. The speculation that
such terms arise from  quantum gravity corrections \cite{Weinberg}
gives the neutrino mass of the range of  $10^{-5} eV$, which is much
less than the experimental estimations.

In the formulation of noncommutative geometry an important role is played by a finite spectral
triple, that is a geometry over a finite algebra. We shall not assume any particular model
here, taking the resulting generalized Dirac operator and the Higgs potential for granted.

\section{The extended spectral action principle}

In the model-building principle of noncommutative geometry, the main role is played
by an algebra, which corresponds to the functions on the space-time, its representation
on the Hilbert space and the Dirac operator, which encodes all the information about
differentiation, metric as well all internal degrees of freedom. Dirac operator could be
modified by its {\em internal fluctuations} thus leading to a family of operators.

The physics is set by the action (\ref{boso}) with a suitable cutoff function $f$. The asymptotic
expansion of the bosonic action leads to leading terms providing the volume (cosmological constant),
the traditional Einstein-Hilbert action of pure gravity as well as the Yang-Mills gauge functionals
for the internal fluctuations of the Dirac operator, which are identified as gauge fields. Clearly, one
has to make a Wick rotation to Minkowski geometry to consider the physical fields, nevertheless
the consistency with the standard physical picture is striking.

As it has been already observed by Chamseddine \cite{Chams} there is
a huge difference between the bosonic and fermionic (\ref{ferm})
parts of the action in Noncommutative Geometry, especially in the
formulation of the spectral action principle. At first, the bosonic
(and gravitational) part of the action depends solely on the
eigenvalues (or, more precisely, on the eigenvalue asymptotic) of
the generalized Dirac operator. This is not the case for fermion
fields, where the action principle is the expectation value of the
Dirac operator in the state set by $\Psi$. Thus all the eigenvalues
of the Dirac do intervene. The generalization proposed by
\cite{Chams} and tested on a simple example led to the
interpretation of additional terms as arising from the
supersymmetric theory.

Although his efforts were concentrated on the couplings between gauge fields and fermions there
is still a place in such models for couplings between discrete gauge field strength and fermions,
in particular, for coupling between two fermions and a term quadratic in the Higgs. The spectral
approach to the action in noncommutative geometry offers a feasible theoretical mechanism for the
appearance of such terms.

First of all, one might consider a {\em total} action of the type:
\begin{equation}
S = \hbox{Tr\ } f \left( (D + P_\Psi)^2 \right),
\label{ferm1}
\end{equation}
where $f$ is some cutoff function and $P_\Psi$ is a projection on the field $\Psi$.

To give an example what are the consequences, consider the terms in the asymptotic expansion
that shall arise. Using standard results of Gilkey \cite{Gilkey} for the asymptotic expansion,
we obtain that the relevant terms involving fermions read:
\begin{equation}
\begin{aligned}
S_\Psi  =& \Lambda^2 \int_M \hbox{Tr\ } (D P_\Psi + P_\Psi D) +
\frac{1}{360} \left( \int_M -60 R\, \hbox{Tr\ } D P_\Psi \right. \\
&+ \left. 180\, \hbox{Tr\ } (D P_\Psi + P_\Psi D)^2 + 60 \hbox{Tr\ } \triangle (D P_\Psi) \right).
\end{aligned}
\end{equation}

The leading term in the expansion is nothing else than the standard fermion action:
$$ \int_M <\Psi, D \Psi>, $$
which gives the minimal couplings between fermion and gauge fields, as well as coupling
between fermions and the Higgs and then fermion masses.

The terms of the next order contain some higher-derivative components, coupling
of fermions to the scalar curvature as well as a nonlinear coupling of the fermions
to the Dirac operator. Let us analyze the latter in more details. Simple calculation
yields that the next-order terms that do not involve derivatives are:
\begin{equation}
\hbox{Tr\ } (D P_\Psi + P_\Psi D)^2 = 3 (<\Psi | D \Psi>)^2 + <\Psi | D^2 \Psi>.
\label{nextterms}
\end{equation}

{}From our point of view, the interesting terms are these, which include the
square of the Dirac operator.

\section{The Ansatz}

Let us postulate a simple-minded solution to the question "how to obtain quadratic terms"
(\ref{inter}) using an extension of the spectral action principle.Of course, since the minimal
Standard Model alone does not provide the answer and we must find a way so that no linear term,
giving {\em bare} neutrino mass appears but there will be a quadratic term, originating in a
nonzero contribution (\ref{nextterms}).

First, observe that the expression:
\begin{equation}
\left( H^T \sigma L \right),
\label{inter2}
\end{equation}
where $L$ is a lepton doublet, $H$ is the Higgs doublet and $\sigma$ is a $\sigma^2$ Pauli matrix
is itself gauge-invariant. Recall that under $U(1) \times SU(2)$ gauge transformations the lepton
doublet and the Higgs field transform as follows:
$$  L \mapsto h L \bar{z},  \;\;\;  H \mapsto h H z, $$
with $h \in SU(2), z \in U(1)$. Hence, the term (\ref{inter2}) remains
invariant due to the fact that $\sigma$ intertwines the fundamental representation
of the quaternions with its conjugate:
\begin{equation}
\sigma h \sigma^{-1}= h^* . \label{s2}
\end{equation}
For this reason, any spinor field $N$, which is totally non-interacting shall give
a gauge invariant term of the form:

\begin{equation}
\langle N |  H^T \sigma L \rangle.
\end{equation}

Clearly, adding the field $N$ to the family of all fermion fields is the solution, which is comparable
to the addition of sterile neutrinos.

However, we propose a way so that the effective Standard Model action (as we see it) ignores
those particles. Our {\em Ansatz} is for the spectral action including fermions and the form
of the cutoff function $f$ that excludes a subspace of the Hilbert space.

We assume that the full Hilbert space includes some sterile, noninteracting (that is, one
which is not in the representation of the discrete algebra) fermion $N$. Furthermore, instead of taking
$f$ to be scalar-valued let us assume that in addition to the cutoff in the eigenvalues of $D$, the
function projects on a subspace of the Hilbert space, which consists of all fermions, which are in the
representation of the discrete algebra. One can explain this restriction saying that the spectral action
presents an effective action up to a some fixed energy scale and the cut-off is implemented also by the
restriction to some subspace of the Hilbert space.

If we look now at the spectral action:

\begin{equation}
S_{eff} = \hbox{Tr}_{ph} \, f_\Lambda((D + P_\Psi)^2),
\end{equation}

we see that neither the bosonic part nor the standard fermionic part (that is linear
in $D$) shall change with respect to the well-known action of the Standard Model.
The difference shall appear, however, at the level of correction terms, that is,
second-order with respect to the leading term. There, we shall see contributions
from $D^2$, which are exactly of the form:
$$ \sigma H^* H^T \sigma, $$

Even though the details of the effective action depend on the particular form of $f$, it is not
important for our considerations. The modified action provides just corrections to the original
action (\ref{ferm}) of order $o( \frac{1}{\Lambda^2})$.

If we take the value of the cutoff parameter \cite{CoCha} $\Lambda = 10^{15} GeV$ and estimate the
resulting neutrino mass (taking the coefficient in the term to be of order $1$) we obtain the values
of order $10^{-2} eV$, which agrees with the current experimental data. In fact, it was pointed out
that such order of neutrino mass suggests in turn \cite{Wilczek} that the scale of $10^{15} GeV$
(which is well below the Planck scale) is the one at which one can expect to change the effective
model.

\section{Conclusions}

We have shown that a minimal adjustment of the contents of the
Standard Model together with the extension of the spectral action
principle to fermions provides an explanation of the small neutrino
mass. Clearly, the correction terms have no measurable influence on
the masses of other particles as they are many orders of magnitude
smaller. In the case of originally massless neutrino this correction
shall be, however, a leading term. Therefore, even within the
simplest description of the Standard Model in noncommutative
geometry it is possible to generate a neutrino mass via correction
from the spectral action. We presented argument for a one family but
its generalization to many generations is straightforward. Of
course, the terms are not renormalizable. However, we might treat
the model as an effective one and the neutrino mass term as the
effective at a given energy scale. Incidentally, the extra gravity
terms that appear in the spectral action principle as next-order
corrections to the Einstein-Hilbert action lead to similar problems.

If we take the noncommutative geometry description and the spectral
action as an approximation of the real geometry, then the cutoff
parameter $\Lambda$ of the bosonic action has a natural physical
meaning of the scale of possible fluctuations of the geometry. It is
conceivable that the physics observed at today energy scales is the
restriction of some more general model. The spectral action appears
to be well-suited for this purpose.

The proposition shown in this paper indicates a way of introducing the neutrino mass in a purely
dynamical way. Although the term (\ref{inter}) that induces this mass at the nonzero Higgs expectation
value is not new, the method of obtaining it is set into the noncommutative geometry. Since the part
of the Hilbert space, which is cut-off through the spectral action corresponds to some hypothetical
particles of the sterile neutrino type - one might view it as a form of see-saw mechanism \cite{seesaw}.
Although no direct mass of the extra particles is needed - the cut-off mechanism allows for the appearance
of correction terms, which are significant only in the case of originally massless neutrinos.

Note that contrary to the see-saw mechanism we do not need to justify the non-dynamical and large
mass terms for the sterile neutrinos, as in our case the role of the "see-saw weight" is played by
the cut-off parameter. We believe that the possibility appears to be feasible and requires more
research on the action principle for fermions and quantum theory of fields in this noncommutative setup.
Finally, let us mention that since the presented model keeps the postulate that there are only left neutrino
currents, the experimental results (in particular, from neutrinoless double $\beta$ decay) might distinguish
whether the "discrete" noncommutative geometry description of the Standard  Model is right or wrong. We believe
that the geometric approach, that we advocate here, might also shed a new light on the issues of physics
beyond the Standard Model.

{\bf Acknowledgements:}
The author would like to thank the Hausdorff Research Institute  for Mathematics
for hospitality and support.

\end{document}